\documentclass[prl,amsmath,amssymb,twocolumn, showpacs, superscriptaddress,10pt]{revtex4}

\usepackage[english]{babel}
\usepackage[utf8]{inputenc}
\usepackage[T1]{fontenc}
\usepackage{amsmath}
\usepackage{hyperref}
\usepackage{graphicx}
\usepackage{amsfonts}
\usepackage{amsthm}

\theoremstyle{remark}

\usepackage{color}
\definecolor{Blue}{rgb}{0.00, 0.00, 1.00}
\definecolor{Red}{rgb}{1.00, 0.00, 0.00}

\newcommand{\nn}{\nonumber}
\newcommand{\be}{\begin{equation}}
\newcommand{\ee}{\end{equation}}
\newcommand{\bea}{\begin{eqnarray}}
\newcommand{\eea}{\end{eqnarray}}

\newcommand{\beq}{\begin{equation}}
\newcommand{\eeq}{\end{equation}}
\newcommand{\beqn}{\begin{eqnarray}}
\newcommand{\eeqn}{\end{eqnarray}}

\def\XXint#1#2#3{{\setbox0=\hbox{$#1{#2#3}{\int}$}
     \vcenter{\hbox{$#2#3$}}\kern-.5\wd0}}

\begin{document}

\setlength{\abovedisplayskip}{5pt}
\setlength{\belowdisplayskip}{5pt}

\title{Statistics of extremes in eigenvalue-counting staircases}

\author{Yan V. Fyodorov}
\affiliation{King's College London, Department of Mathematics, London  WC2R 2LS, United Kingdom}
\affiliation{L.D.Landau Institute for Theoretical Physics, Semenova 1a, 142432 Chernogolovka, Russia}
\author{Pierre Le Doussal}
\affiliation{Laboratoire de Physique Th\'eorique de l'Ecole Normale Sup\'erieure,
PSL Research University, CNRS, Sorbonne Universit\'es,
24 rue Lhomond, 75231 Paris, France}

\begin{abstract}
We consider the number ${\cal N}_{\theta_A}(\theta)$ of eigenvalues $e^{i \theta_j}$
of a random unitary matrix, drawn from CUE$_{\beta}(N)$,
in the interval $\theta_j \in [\theta_A,\theta]$. The deviations from its mean,
${\cal N}_{\theta_A}(\theta) - \mathbb{E}({\cal N}_{\theta_A}(\theta))$, form a random process
as function of $\theta$. We study the maximum of this process,
by exploiting the mapping onto the statistical mechanics of log-correlated
random landscapes. By using an extended Fisher-Hartwig conjecture 
supplemented with the freezing duality conjecture for log-correlated fields, we obtain the cumulants of
the distribution of that maximum for any $\beta>0$.  It exhibits combined features of standard
counting statistics of fermions (free for $\beta=2$ and with Sutherland-type interaction
for $\beta\ne 2$)  in an interval and extremal statistics of the fractional Brownian motion
with Hurst index $H=0$. The $\beta=2$ results are expected to apply to
the statistics of zeroes of the Riemann Zeta function.
\end{abstract}



\maketitle

\newpage

Characterizing the full counting statistics of the fluctuations of the number ${\cal N}$ of $1d$ fermions
in an interval is important in numerous physical contexts, both for ground state and dynamical properties.
It appears e.g. in shot noise \cite{Levitov}, in fermion chains \cite{AbanovIvanovQian2011,IvanovAbanov2013}, in interacting Bose gases \cite{Calabrese_etal}, in non-equilibrium Luttinger liquids \cite{PGM}, in trapped fermions \cite{Eisler1,MarinoVariance,DeanPLDReview}, and for studying related observables,
such as the entanglement entropy \cite{KeatMezzadri,Caux2019,CalabresePLDEntropy} or the statistics of local magnetization in quantum spin chains \cite{EiserRacz2013}.
An equivalent problem can be formulated as counting eigenvalues
of large random matrices (RM). As is well known since Dyson's work \cite{Dyson}, such eigenvalues
behave as classical particles with 1-d Coulomb repulsion at inverse temperature $\beta>0$.
Namely, consider a unitary $N \times N$ matrix $U$ and denote the corresponding unimodular eigenvalues as $z_j=e^{i \theta_j}$, $j=1,\ldots, N$, with phases $\theta_i \in ]-\pi,\pi]$.  Then for any given $\beta>0$ one can construct the so-called Circular $\beta$-Ensemble CUE$_{\beta}(N)$ in such a way that the
expectation of a function depending only on the eigenvalues of $U$ will be given by
\be \label{1}
\mathbb{E}(F) =
c_N \prod_{j=1}^N \int_{-\pi}^{\pi} d\theta_i \prod_{1 \leq j < k \leq N}
|e^{i \theta_j} - e^{i \theta_k}|^{\beta}\, F
\ee
where $F\equiv F(\theta_1,\dots,\theta_n)$.
For $\beta=2$ such matrices can be thought of as drawn uniformly according to the corresponding Haar's measure on $U(N)$,  whereas for a generic $\beta>0$ the explicit construction is more involved, see \cite{Killip}.
 For any $\beta>1$, the r.h.s of \eqref{1} equals the quantum expectation value
of $F$ in the ground state of $N$ spinless fermions, of coordinates
$\theta_i$ on the unit circle,  described by the Sutherland Hamiltonian
\cite{Sutherland}  $H = - \sum_i \frac{\partial^2}{\partial \theta_i^2} + \sum_{i<j}
\frac{\beta(\beta-2)}{8 \sin^2\left(\frac{\theta_i-\theta_j}{2}\right)}$. For $\beta=2$, Eq \eqref{1} thus describes
non-interacting fermions, while for $\beta \neq 2$ the fermions interact, via an
inverse square distance pairwise potential.

\begin{figure}[t]
        \includegraphics[width=0.45\textwidth,height=\textheight,keepaspectratio=true]{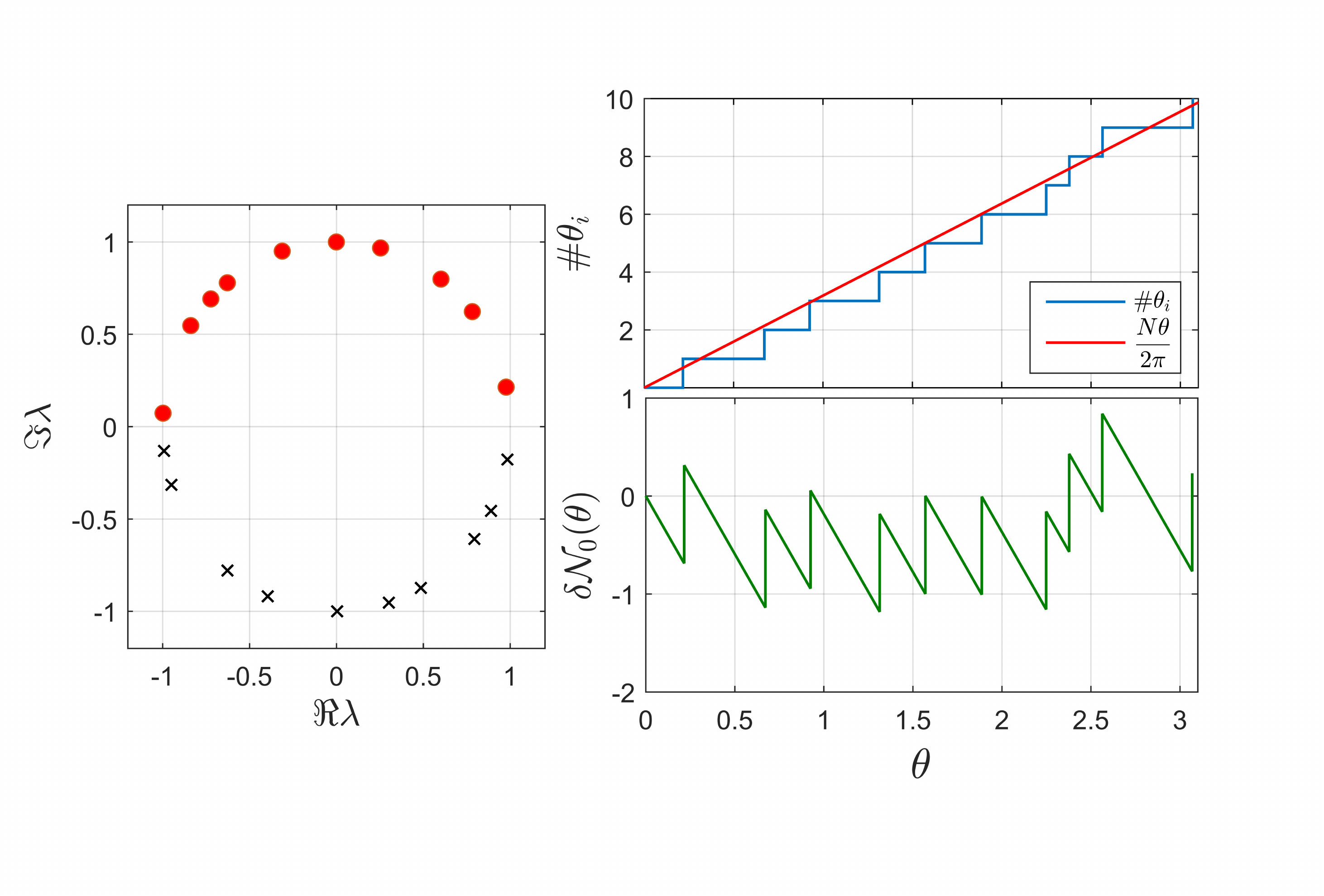}
        \caption{\small Constructing an instance of $\delta N_0(\theta)$ for
        $\theta\in [0,\pi]$ for $\beta=2$ and $N=20$.  Left: eigenvalues $\lambda=e^{i\theta_i}$. Right: counting staircase
        (top), with mean subtracted (bottom).}
        \label{fig:hinv1}
    \end{figure}

\begin{figure}[t]
        \includegraphics[width=0.25\textwidth,height=\textheight,keepaspectratio=true]
        {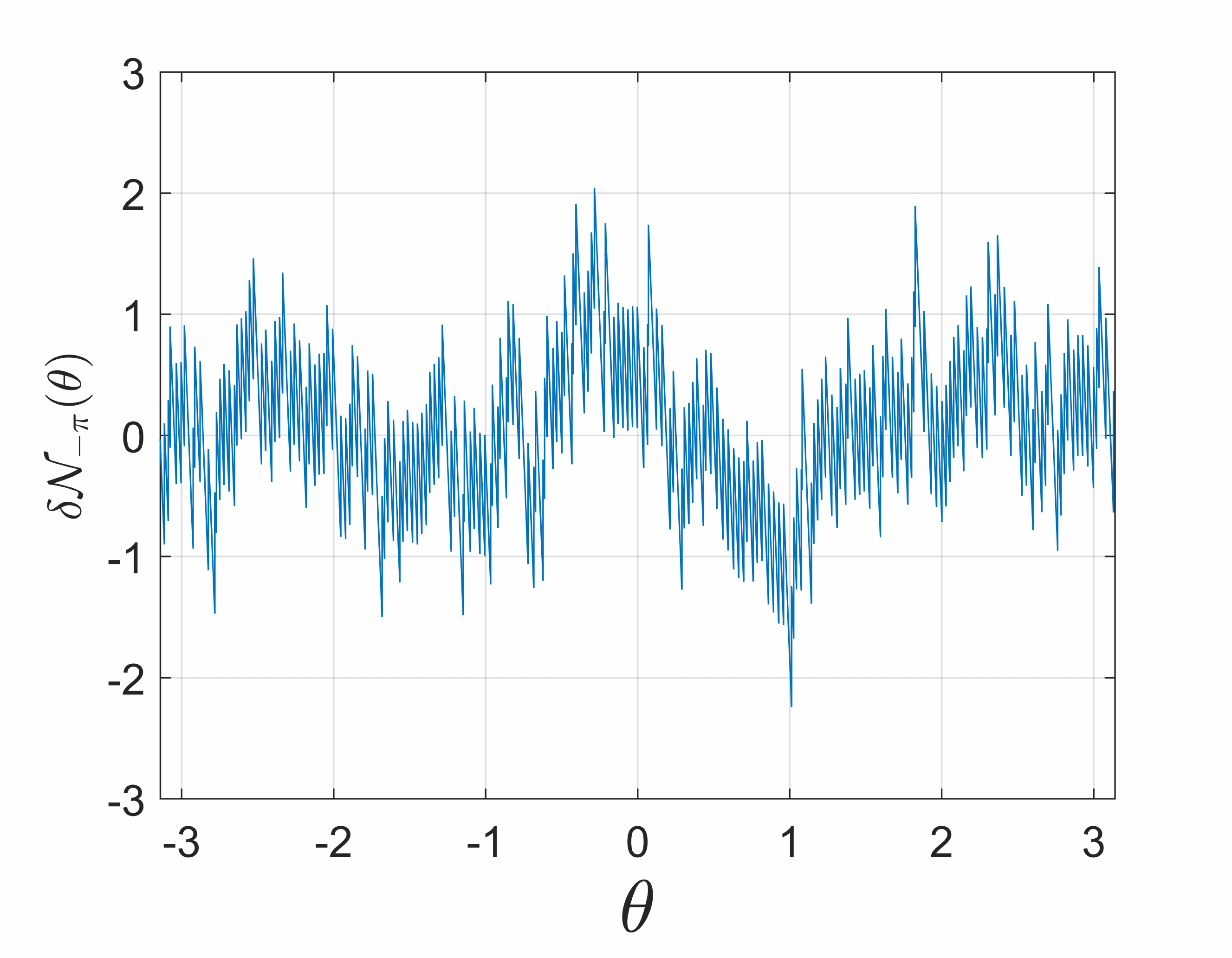}
        \caption{\small A single realization of $\delta N_{-\pi}(\theta)$ for the full circle
        $\theta\in [-\pi,\pi]$ for $\beta=2$ and $N=200$.}
        \label{fig:hinv2}
    \end{figure}

Let us now define the number of eigenvalues/fermions, ${\cal N}_{\theta_A}(\theta)$,
in the interval $\theta_j \in [\theta_A,\theta]$ as
\be\label{defcount}
{\cal N}_{\theta_A}(\theta) = \sum_{j=1}^N \left(\chi(\theta - \theta_j) - \chi(\theta_A-\theta_j) \right)
~ , ~  \chi(u) = \begin{cases} 1 ~ ,~ u>0 \\
0 ~ , ~ u<0 \end{cases}
\ee
As a function of $\theta$ this is a staircase with unit jumps upwards at random
positions {$\theta_j\in [\theta_A,\theta]$}. The mean slope (i.e. the mean density of {eigenvalues}/fermions) being constant, the mean profile is $\mathbb{E}({\cal N}_{\theta_A}(\theta)) = \frac{N (\theta-\theta_A)}{2 \pi}$. In a given {random matrix realization}/sample one
can define the deviation to the mean, $\delta {\cal N}_{\theta_A}(\theta)={\cal N}_{\theta_A}(\theta)  - \mathbb{E}({\cal N}_{\theta_A}(\theta) )$, and study it as a random process as a function of $\theta$,
i.e. as a function of the length of the interval $\theta-\theta_A$,  see Fig. \ref{fig:hinv1}
and \ref{fig:hinv2}. From the view of such a process, the standard
results on fermion counting statistics \cite{AbanovIvanovQian2011},
encoding the full distribution of $\delta {\cal N}_{\theta_A}(\theta)$ for a {\it fixed}
value of $\theta$, is a very local information. Such information is clearly insufficient for
understanding various {\it non-local} properties of the process, such as characterizing {\it maximal deviation} of the staircase from its mean, i.e. $\max_{\theta \in [\theta_A,\theta_B]} | {\cal N}_{\theta_A}(\theta)  - \mathbb{E}({\cal N}_{\theta_A}(\theta) ) |$. After 
normalization this is 
the {\it Kolmogorov-Smirnov} (KS) statistics, an 
outstanding open problem for spectra of random matrices \cite{BaoHe}, \cite{Clayes_etal}.

%
In this Letter we study the value distribution separately for the maximum (and equivalently the minimum) of the centered process 
by explicitly {calculating} the cumulants of the
probability density function (PDF) for the maximum value defined as
\be \label{max}
\delta {\cal N}_m = \max_{\theta \in [\theta_A,\theta_B]} \left\{ {\cal N}_{\theta_A}(\theta)  - \mathbb{E}({\cal N}_{\theta_A}(\theta) ) \right\}
\ee
on an interval $[\theta_A,\theta_B] \subset ]-\pi,\pi]$, of a fixed length $\ell=\theta_B-\theta_A$.
To derive the PDF of $\delta {\cal N}_m$ in the limit $N~\gg~1$ we will show that for scales larger than $1/N$ the process $\delta {\cal N}_{\theta_A}(\theta)$ is very close to a special  version of $1D$ log-correlated Gaussian field, the so called fractional Brownian Motion with Hurst index $H=0$, denoted as fBm0, defined in \cite{FyoKhorSimm} and whose extrema where investigated recently \cite{FLD2016,CaoPathologies}.
However it turns out that the relation to fBm0 alone is insufficient to fully
determine the statistics of $\delta {\cal N}_m$.
Namely, we will demonstrate that although the process $\delta {\cal N}_{\theta_A}(\theta)$
for large $N\gg 1$ is very close to the fBm0 at {\it different} points,
the {\it non-Gaussian} features which characterize its {\it single-point} statistics
show up in a non-trivial way in the PDF of its maximum $\delta {\cal N}_m$.
These single-point features are inherited from the discrete nature of the number
of fermions/eigenvalues as exemplified e.g. in  fermion counting statistics \cite{AbanovIvanovQian2011}.


We now describe our main findings {by first
assuming} that the Dyson parameter is rational
and can be represented as $\beta/2=s/r$ where $s$ and $r$ are mutually prime, {and relaxing this assumption
later on}. We find that, for any fixed interval, the mean value of the maximum $\delta {\cal N}_m$  defined in \eqref{max} exhibits, for $N \to \infty$, the universal behavior of the log-correlated fields \cite{DS1988,CLD,FLDR2,DRZ} :
\be \label{first}
2 \pi \sqrt{\frac{\beta}{2}} \mathbb{E}(\delta {\cal N}_m) \simeq 2 \log N - \frac{3}{2} \log \log N + c^{(\beta)}_\ell
\ee
where $c^{(\beta)}_\ell=O(1)$ is an unknown $\ell$-dependent constant.
The variance for the maximum $\delta {\cal N}_m$ exhibits to the leading order the extensive universal logarithmic growth typical for {\it pinned} log-correlated fields \cite{CaoPathologies}, on top of which we can evaluate the corrections of the order of unity:
\be \label{second}
 \mathbb{E}^c (\delta {\cal N}_m^2) \simeq  \frac{2}{\beta (2 \pi)^2}  (2 \log N + \tilde C^{(\beta)}_2 + C_2(\ell))
\ee
Finally, the higher cumulants converge to a finite limit as $N\to \infty$:
\be \label{k}
 \mathbb{E}^c (\delta {\cal N}_m^k) \simeq  \frac{2^{k/2}}{\beta^{k/2} (2 \pi)^k} (\tilde C^{(\beta)}_k + C_k(\ell)),
\ee
where the constants $C_k(\ell)=O(1)$ depend on the length $\ell$ of the interval and
will be given below in two limiting cases. The $\ell-$independent constants $\tilde C^{(\beta)}_k$ for $k \geq 2$ are given by
\be \label{Cbeta}
\tilde C^{(\beta)}_k =  \frac{d^k}{dt^k}|_{t=0} \log ( A_{\beta}(t) A_{\beta}(-t) )
\ee
where
\bea \label{AB}
&& A_{\beta}(t) =r^{-t^2/2} \prod_{\nu=0}^{r-1} \prod_{p=0}^{s-1}\frac{G(1-\frac{p}{s}+\frac{\nu+i t\sqrt{\frac{2}{\beta}}}{r})}
{G\left(1-\frac{p}{s}+\frac{\nu}{r}\right)}
\eea

Here $G(z)$ denotes the standard Barnes function satisfying $G(z+1)=\Gamma(z) G(z)$, with $G(1)=1$. Note that all the odd coefficients
$\tilde C^{(\beta)}_{2k+1}$ vanish. Specifying for $\beta=2$, one has $A_2(t)=G(1+ i t)$, leading to $\tilde C^{(2)}_2 = 2(1+ \gamma_E)$ and $\tilde C^{(2)}_4= -12 \zeta (3)$.
 {Notably, using \eqref{Cbeta}, \eqref{AB},
we were able} to obtain a formula for the $\tilde{C}^{(\beta)}_k$ as single infinite series \cite{SM}, which
shows that they are smooth as a function of the Dyson parameter $\beta$, thus relaxing the assumption of rationality. As discussed below, the factors $ A_{\beta}(t)$, hence $\tilde{C}^{(\beta)}_k$, are intimately but non-trivially related to the cumulants
of the number of fermions (free for $\beta=2$ and with Sutherland-type interaction
for $\beta\ne 2$) in a mesoscopic interval of the circle.

By contrast the factors $C_k(\ell)$  are $\beta$-independent and originate from the problem of
the maximum of a fBm0  on the interval $[\theta_A,\theta_B]$.
For the $\ell$-dependent constants we obtain explicit formula in two cases:

{\it (i) maximum over the full circle $\ell=2 \pi$}.
In that case $[\theta_A,\theta_B]=]-\pi,\pi]$ and we find for any $k \geq 2$
\bea \label{rescircle}
C_k(2 \pi) = (-1)^k \frac{d^k}{dt^k}|_{t=0}  \log \bigg[
 \frac{\Gamma(1+ t )^2 G(2-2 t)}{G(2-t)^3 G(2+t)} \bigg]
\eea
which is related to the fBm0 bridge on $]-\pi,\pi]$
studied in \cite{CaoPathologies}

{\it (ii) maximum over a mesoscopic interval $\frac{1}{N} \ll \ell \ll 1$}.

For $k \geq 2$ we obtain in this regime
\bea \label{resinterval}
&& C_k(\ell) \simeq 2 \log \ell \, \delta_{k,2} \\
&& + (-1)^k \frac{d^k}{dt^k}|_{t=0} \bigg[
\frac{2  \Gamma(1+ t )^2  G(2-2 t)}{G(2+t)^2  G(2-t)
G(4-t)} \bigg]
\eea
This result is related to the fBm0 on an interval,
with one pinned and one free end,
studied in \cite{CaoPathologies}.
Note that the variance depends logarithmically on $\ell$
at small $\ell$, whereas higher cumulants have limits as
$\ell \to 0$.  Note that $l\to 0$ limit is expected to provide
 the $L\gg 1$ asymptotic for statistics of the maximum of ${\cal N}_{\theta_A}(\theta)$ in intervals of the order $2\pi L/N$, comparable with the mean eigenvalue spacing. The universal statistics of
CUE$_{\beta}$ eigenvalues at such local scales is described by the so called {\it sine}-$\beta$ process  \cite{SineBetaVirag}
and the associated counting function has been studied in \cite{DH19}.

Finally, addressing the question of the {\it location} of the maximum in \eqref{max}, $\theta_m \in [\theta_A,\theta_B]$, let us define
$y_m = (\theta_m - \theta_A)/\ell$. For the mesoscopic interval, we predict the PDF of $y_m$
to be symmetric around $\frac{1}{2}$, with
$\mathbb{E}(y_m^2)= \frac{17}{50}$ and
$\mathbb{E}(y_m^4)= \frac{311}{1470}$, thus
deviating from the uniform distribution.  For the full circle we
find a uniform distribution for $\theta_m$ \cite{footnote2}.
However, joint moments for the position {\it and} value
of the maximum 
show the
effect of pinning at $\theta=\theta_A$ (see details in \cite{SM}).


To elucidate the relation to fBm0,
let us recall that
the process $\delta N_{\theta_A}(\theta)$ is exactly given by the difference \cite{SM}
\be \label{eq5}
\delta {\cal N}_{\theta_A}(\theta) = \frac{1}{\pi} {\rm Im} \log \xi_N(\theta) -
 \frac{1}{\pi} {\rm Im} \log \xi_N(\theta_A)
\ee
where $\xi_N(\theta)= \det ( 1 - e^{- i \theta} U)$ is the characteristic polynomial (CP). 
As 
shown in \cite{HKOC} for $\beta=2$ (see \cite{CMN} for general $\beta>0$) the joint probability density of ${\rm Im} \log \xi_N(\theta)$ at two 
distinct points $\theta_1\ne \theta_2$ converges as $N \to +\infty$
to that of a Gaussian  process $W_{\beta}(\theta)$ of zero mean and covariance
\be \label{covcharpol}
\mathbb{E}( W_{\beta}(\theta_1) W_{\beta}(\theta_2))  = - \frac{1}{2\beta} \log \left[ 4 \sin^2\left(\frac{\theta_1-\theta_2}{2}\right) \right]
\ee
a particular instance of the 1D log-correlated Gaussian field.
 Since \eqref{eq5} implies that $\delta {\cal N}_{\theta_A}(\theta=\theta_A)=0$ in any realization, the relevant object is the 
{\it pinned log-correlated process} closely related to fBm0.  The log-correlated fields being highly singular always require a regularization to study their value distribution.
The imaginary parts of the $\log \xi_N(\theta)$ for $N \gg 1$
provides such a natural regularization \cite{CMN,Webb,BWW,Lambert2019}, being asymptotically a random process $W$
which shares the covariance (\ref{covcharpol}) but 
with 
a finite variance $\mathbb{E}( W(\theta)^2)=\beta^{-1}\log{N}+O(1)$. 
Via (\ref{eq5}) this 
provides the well-known asymptotic of the eigenvalues/fermions number variance: $\mathbb{E} (\delta {\cal N}^2(\theta)) \simeq  \frac{2}{\beta \pi^2}  \log N$. We shall see however \cite{SM} that naively replacing the difference $\delta {\cal N}_{\theta_A}(\theta)$ with its Gaussian approximation $\frac{1}{\pi}\left[ W_{\beta}(\theta)-W_{\beta}(\theta_A)\right]$ (
related to the 
{\it bosonization} of the fermionic problem)  is not sufficient for 
characterizing the maximum
of the process.

Gaussian fields 
characterized by a logarithmic covariance 
appear in 
chaos and turbulence \cite{Kahane},
branching random walks and polymers on trees~\cite{DS1988,CLD},
multifractal disordered systems 
~\cite{CMW,logmultlec},
two-dimensional gravity~\cite{RVLectures,KRV-DOZZ}.
Early works on their {\it extrema}
revealed a connection to a remarkable freezing transition \cite{DS1988,CMW,CLD}. Through exact solutions, it
led to predictions for the PDF of the maximum value of a log-correlated field
on the circle and on the interval ~\cite{FB2008,FLDR},
involving the 
{\it freezing duality conjecture} (FDC) (see \cite{FLD2016}
for an extensive discussion). 
This led to 
further results in theoretical and mathematical physics
~\cite{FLDR2,Ostr2016,OstrRev,Cao2ndMin,CaoLiouville,CaoOPELiouville} and probability
~\cite{MadauleFreezingProof,BerLectures,DRZ,SubagZeitouni,FyoSimm,ABB,PZ2017,Remy1,Remy2,BL}.
A log-correlated context of random CP
attracting a lot of attention ~\cite{FHK_prl,FyoKeat2014,ABH,ABBRS,Najnudel,AOR,Harper, FyoSimm,ABB,PZ2017,CMN,Najnudel2}, none of these studies yet addressed the eigenvalue/zeros counting function in the intervals $\ell=O(1)$.



%

To study the maximum of the random field
$\delta {\cal N}(\theta)$ we follow 
\cite{FB2008,FLDR,FLD2016,FHK_prl,FyoKeat2014}
and introduce a statistical mechanics problem of
partition sum:
\be \label{Zb}
Z_{b} = \frac{N}{2 \pi} \int_{\theta_A}^{\theta_B} d \phi
\, e^{2 \pi b \sqrt{\beta/2}\, \delta {\cal N}_{\theta_A}(\phi)},
\ee
The `` inverse temperature" 
is equal to
$-2 \pi b \sqrt{\beta/2}$, and we choose $b>0$ since we are
studying here the maximum 
retrieved from the free energy ${\cal F}$ for $b\to +\infty$ as
\be \label{limit}
\delta {\cal N}_m = \lim_{b \to +\infty} {\cal F}
\quad , \quad
{\cal F} =  \frac{1}{2 \pi b \sqrt{\beta/2}} \log Z_{b}
\ee
To study the statistics of the associated free energy we start with
considering the integer moments of $Z_b$ given by
\bea
&& \! \! \mathbb{E}[Z_b^n]  = 
 \left(\frac{N}{2 \pi}\right)^n  \int_{\theta_A}^{\theta_B}
e^{- b \sqrt{\beta/2} \sum_{a=1}^n N (\phi_a-\theta_A)}\, \prod_{a=1}^n d \phi_a \nn \\
&& \times
\, \mathbb{E} [ \prod_{j=1}^N e^{2 \pi b \sqrt{\beta/2} \sum_{a=1}^n
(\chi(\phi_a-\theta_j) - \chi(\theta_A-\theta_j) )}] \label{15}
\eea
The expectation value in \eqref{15} over the CUE$_{\beta}(N)$ computed using \eqref{1}
has the form $\mathbb{E}[ \prod_{j=1}^N g(\theta_j) ]$
where we defined
\be
\log g(\theta)=  2 \pi b \sqrt{\beta/2} \sum_{a=1}^n
(\chi(\phi_a-\theta) - \chi(\theta_A-\theta) )
\ee
This can be further rewritten for any $\phi_a,\theta,\theta_A \in ]-\pi,\pi]$ with $\phi_a > \theta_A$
as
\bea \label{symb}
&& \log g(\theta)=  b \sqrt{\beta/2}
[  \sum_{a=1}^n \phi_a- n \theta_A  \\
&& +  n \arg e^{i (\theta_A - \theta + \pi)}
-  \sum_{a=1}^n \arg e^{i (\phi_a - \theta + \pi)} ] \nn
\eea
where we define the arg function as
\be \label{jump}
{\rm arg} e^{i \phi} = \begin{cases}
\phi \quad -\pi < \phi \leq \pi \\
\phi- 2 \pi \quad \pi < \phi \leq 3 \pi
\end{cases}
\ee
  For $\beta=2$,
$\mathbb{E}[ \prod_{j=1}^N g(\theta_j) ]=\det_{1 \leq j,k \leq N}[g_{j-k}]$ is a Toeplitz determinant,
where $g_p=\int_{-\pi}^{\pi} \frac{d\theta}{2 \pi}  e^{-i p \theta} g(\theta)$ is the associated symbol, and $g(\theta)$
according to (\ref{symb})-(\ref{jump}) has jump singularities. The corresponding asymptotics as $N\to \infty$
is given by the famous Fisher-Hartwig (FH) formula \cite{FH1969}  proved rigorously in \cite{DeiftItsKrasovsky2009}.
For a general rational $\beta$ extension of FH formula has been conjectured in \cite{FF2004}.
Specifying the expressions
in \cite{FF2004} to our case gives for $N \to +\infty$ and $n b^2<1$\\
\bea \label{Zn}
&& \mathbb{E}[Z_b^n]  \simeq
\left(\frac{N}{2 \pi}\right)^n  N^{b^2 (n + n^2) }
|A_{\beta}(b)|^{2n} |A_{\beta}(b n)|^{2} \nn \\
&& \times \int_{\theta_A}^{\theta_B} \label{momasy}
\prod_{1 \leq a < c \leq n}
|1- e^{i (\phi_a-\phi_c)}|^{- 2 b^2} \\
&&
\times \prod_{1 \leq a \leq n}
|1- e^{i (\phi_a- \theta_A)}|^{2 n b^2}\, \prod_{a=1}^n d \phi_a \nn
\eea
where the function $A_{\beta}(b)$ is defined in
\eqref{AB}. 
Had we used instead 
an approximation replacing the
difference $\delta {\cal N}_{\theta_A}(\theta)$ in the large$-N$ limit with the logarithmically correlated Gaussian process $W_{\beta}(\theta)$ defined via (\ref{eq5}) - (\ref{covcharpol}),
{we would reproduce the Coulomb gas factors in (\ref{Zn}) but}
miss the 
factors $A_{\beta}(b)$, see \cite{SM}. Hence, this product  encapsulates the residual non-Gaussianity of the 
process.

Let us first discuss the simplest case $n=1$
when \eqref{Zn}  can be interpreted, via (\ref{Zb}), as giving
\be \label{onepoint}
 \mathbb{E}(e^{2\pi b \delta {\cal N}_{\theta_A}(\theta)})  \simeq
N^{2 b^2  } |A_{\beta}(b )|^{4} \left(4 \sin^2 \frac{\theta- \theta_A}{2}\right)^{b^2}
\ee
This formula 
can be interpreted as the generating
function for the full counting statistics for the number of Sutherland-model fermions in an
interval of size $\theta- \theta_A$ which seems not to be addressed in the literature apart from
 the free-fermion case $\beta=2$ \cite{AbanovIvanovQian2011,footnote1}  and $\beta=4$ \cite{beta4}.

Further progress is possible in the two cases when the Coulomb integrals in
\eqref{Zn} can be explicitly calculated.

{\it (i) Full circle $\theta_A = - \pi$, $\theta_B=\pi$}.
In that case the Coulomb integral is known as
the Morris integral \cite{ForresterWarnaar} leading to
\bea
 \mathbb{E}[Z_b^n]  &\simeq&
\left(\frac{N}{2 \pi}\right)^n N^{b^2 (n + n^2) }
|A_{\beta}(b)|^{2n} |A_{\beta}(b n)|^{2}  \nn \\
& \times&  {\bf M}(n,a=-n b,b)
\eea
where ${\bf M}(n,a,b)$ is defined  Eq (14) in \cite{CaoPathologies}.
This result is valid in the high temperature phase with $n b^2<1$.
From this expression for integer moments there is a well defined
procedure to obtain  the double sided Laplace transform (DSLT) of the
free energy first in the high temperature phase $b<1$ via an analytic continuation.
Defining
$t = - b n$ we obtain
\bea
&& \mathbb{E} \left(e^{- 2 \pi\sqrt{\frac{\beta}{2}} \left({\cal F}-{\cal F}_1\right) t} \right) \simeq N^{- t Q + t^2}
A_{\beta}(t) A_{\beta}(-t)  \nn
 \\
&& \times \Gamma(1+ t b) \frac{G_b(Q-2 t) G_b(Q)^3}{G_b(Q-t)^3 G_b(Q+t)}  \label{eq1}
\eea
where ${\cal F}_1$ is a constant \cite{footnoteF}
and $Q=b + \frac1{b}$ and $G_b(x)$ is the generalized Barnes function, see Eq. (44)
in \cite{FLDR} and \cite{footnote0}.
We note that if we multiply
both sides of the equation by $\Gamma(1+ \frac{t}{b})$, the right hand side is invariant by duality $b \to 1/b$, since formally $G_b(z)=G_{1/b}(z)$.
According to the FDC \cite{FLDR,FLD2016}
we obtain the 
DSLT in the low temperature phase $b>1$. The result can be written as
\be \label{rel}
\mathbb{E} (e^{- 2 \pi \sqrt{\frac{\beta}{2}}{\cal F} t} ) \Gamma(1+ \frac{t}{b})
= \mathbb{E} (e^{- 2 \pi \sqrt{\frac{\beta}{2}}\delta {\cal N}_m t} )
\ee
where the r.h.s. is our main result, i.e. the DSLT of the PDF of $\delta {\cal N}_m$ for the full circle \cite{footnoteComplex}
\bea  \label{ltcircle}
\mathbb{E} (e^{- 2 \pi \sqrt{\frac{\beta}{2}}\delta {\cal N}_m t} )
 &\simeq & N^{- 2t + t^2}\, e^{c t}
 A_{\beta}(t) A_{\beta}(-t)   \nn \\
&& \times  \frac{\Gamma(1+ t )^2 G(2-2 t)}{G(2-t)^3 G(2+t)} \label{DSLT}
\eea
which, according to \eqref{limit}, is the $b \to +\infty$ limit
of the l.h.s of \eqref{rel}. Here $c= \frac{3}{2} \log \log (N) + c'$
and $c'$ is a constant that we cannot determine by this method.
Expansion of Eq. \eqref{ltcircle} around $t=0$ leads to the large $N$ asymptotics
\eqref{first}-\eqref{k} for the cumulants, together with the predicted
values for the coefficients $\tilde C_k^{(\beta)}$ in \eqref{Cbeta} and $C_k(2 \pi)$
in \eqref{rescircle}. The $C_k(2 \pi)$ equal, up to a factor $(-1)^k$, the cumulants
$C_k$ given in
\cite{CaoPathologies} for the fBm0 bridge, checked against numerics there for $k=2,3,4$.
These coefficients are studied in more details in \cite{SM}.

 {\it (ii) Mesoscopic interval}. A similar calculation gives the maximum over a mesoscopic interval $\frac{1}{N} \ll \ell \ll 2 \pi$.
Relegating the details 
to \cite{SM} we simply quote our second main result,
 the DSLT of the PDF of $\delta {\cal N}_m$ for the
small interval limit of small $\ell \ll 1$:
\bea \label{ltinterval}
&& \mathbb{E} (e^{- 2 \pi \sqrt{\frac{\beta}{2}}\delta {\cal N}_m t} )
\simeq (N \ell)^{- 2t  + t^2}\, e^{c t} A_{\beta}(t) A_{\beta}(-t)   \nn \\
&& \Gamma(1+ t )^2
 \frac{2  G(2-2 t)}{G(2+t)^2  G(2-t)
G(4-t)}
\eea
where $c= \frac{3}{2} \log \log N + c''$. Expansion around $t=0$ leads to the same
coefficients $\tilde C_k^{(\beta)}$ which are thus independent of $\ell$
(as can be seen already from \eqref{15})
and to the result for $C_k(\ell)$ in \eqref{resinterval}, again related to
the ones for the fBm0 on an interval given and numerically checked
in \cite{CaoPathologies}. The structure of the above DSLT's in the
complex plane for $t$ is discussed in \cite{SM}. 


%
In conclusion, we obtained the cumulants of the maximum of the deviation
of the counting function from its mean on an interval, for eigenvalues of random
unitary matrices and for free and interacting fermions on the circle. They inherit features both from
the fBm0 log-correlated field and from the fermionic full counting statistics.
Finally, our result for the distribution of $\delta {\cal N}_m$ provides a first step
to study the Kolmogorov-Smirnov statistics for the counting staircases, which would further require the joint PDF of the maximum and minimum (usually non-trivially correlated  \cite{minmaxlog}).

The results for the mesoscopic interval are expected to be universal
for a broader class of random matrix ensembles, as well as for
fermions on a lattice in the dilute limit \cite{footnote5}. %
Finally, it is natural to conjecture that  for $\beta=2$ universality
extends to describing the statistics of the counting staircases
for the nontrivial zeroes $t_n$ of the Riemann zeta-function $\zeta(1/2+it)$ in mesoscopic intervals of the critical line $ t\in \mathbb{R}$. Such zeroes are known to be extremely faithful to the random matrix statistics when analyzed
in appropriate scales \cite{Odlyzko} underlying a fruitful line of applications of associated CP
to understand ensuing features of $\zeta(1/2+it)$\cite{KS2001,Keating_lec,Bourgade2010,BourgadeKuan,HKOC,CMN}.


\acknowledgments

{\bf Acknowledgments:} We thank X. Cao, 
J.P. Keating and G. Lambert for insightful comments on the early version of this paper,
and S. B. Fedeli for his kind assistance with preparing figures.
YVF thanks the Philippe Meyer Institute for Theoretical Physics
at ENS in Paris. PLD acknowledges support from ANR grant ANR-17-CE30-0027-01 RaMa-TraF.


%

%
\newpage

.

\newpage

\begin{widetext}

\bigskip

\bigskip

\begin{large}
\begin{center}

SUPPLEMENTARY MATERIAL \\

\medskip

{\it Statistics of extremes in eigenvalue-counting staircases} \\

\medskip

Yan V. Fyodorov and P. Le Doussal

\end{center}
\end{large}

\bigskip
%

We provide some additional details for some of the calculations described in the manuscript of the Letter.

\section{Cumulant amplitudes $\tilde C^{(\beta)}_k$ as a function of $\beta$}

Let us recall the formula given in the text for the coefficients $\tilde C^{(\beta)}_k$
which enter in the cumulants of the PDF for $\delta {\cal N}_m$, namely, for
$\beta = 2 s/r$, with $s,r$ mutually prime and $k \geq 2$
\be \label{Cbeta2}
\tilde C^{(\beta)}_k =  \frac{d^k}{dt^k}|_{t=0} \log ( A_{\beta}(t) A_{\beta}(-t) )
\ee
\bea \label{AB2}
&& A_{\beta}(t) =r^{-t^2/2} \prod_{\nu=0}^{r-1} \prod_{p=0}^{s-1}\frac{G(1-\frac{p}{s}+\frac{\nu+i t\sqrt{\frac{2}{\beta}}}{r})}
{G\left(1-\frac{p}{s}+\frac{\nu}{r}\right)}
\eea

To obtain more explicit expressions we use that for $k \geq 2$
\be \label{eqG}
\frac{d^k }{d y^k} \log G(x+ y) |_{y=0} = \phi_k(x) :=  (k-1) \psi^{(k-2)}(x) + (x-1) \psi^{(k-1)}(x) - \delta_{k,2}
\ee
where $\psi^{(k)}(x) = \frac{d^{k+1}}{dx^{k+1}} \log \Gamma(x)$.
Hence, for even $k=2 p$, defining $p=s-q$ we obtain
\be  \label{s1}
\tilde C^{(\beta)}_{2p} = (-1)^p \frac{2}{(r s)^p}
\sum_{\nu=0}^{r-1} \sum_{q=1}^{s} \phi_{2p}(\frac{\nu}{r}+\frac{q}{s}) - 2 \log r \delta_{p,1}
\ee
and we recall that odd cumulants vanish.

Since any real $\beta$ can be reached by a sequence $\beta=2 s_n/r_n$ of arbitrary large $s_n,r_n$ we
can obtain an alternative expression valid for any $\beta$ in terms of a convergent infinite series.
We need to distinguish two cases:\\

{\bf Cumulants $C_{2p}$ with $p \geq 2$}. In that case we see that the large $s,r$ behavior in \eqref{s1} is dominated
by the divergence of $\phi_k(x)$ near $x=0$. We use that
\be
\phi_{2p}(x) = - \frac{(2p-1)!}{x^{2 p}} + O(1)
\ee
One finds for $k=2 p$ with $p \geq 2$
\be  \label{s2}
\tilde C^{(\beta)}_{2p} = (-1)^{p+1} 2 (2 p-1)!
\sum_{\nu=0}^{\infty} \sum_{q=1}^{\infty} \frac{1}{(\nu \sqrt{\frac{\beta}{2}} + q
\sqrt{\frac{2}\beta})^{2 p} }
\ee
One of the sum can be carried out leading to two equivalent "dual" expressions
\be  \label{s3}
 \tilde C^{(\beta)}_{2p} = (-2)^{1-p}
\beta^p \sum_{\nu=0}^{\infty} \psi^{(2 p-1)}(1 + \frac{\beta \nu}{2}) = (-2)^{p+1} \frac{1}{\beta^p}
\sum_{q=1}^{\infty} \psi^{(2 p-1)}(\frac{2 q}{\beta})
\ee
where we have used that $\psi^{(2p-1)}(1)=  (2 p-1)! \zeta(2p)$. The above series are
convergent for $p \geq 2$, since at large $x$ one
has $\psi^{(2 p-1)}(x) \simeq \frac{(2p-2)!}{z^{2p-1}}$. Hence the result is
analytic in $\beta>0$. This asymptotics can be used
to obtain the large $\beta$ expansion
\be  \label{s4}
 \tilde C^{(\beta)}_{2p} = (-2)^{1-p}
  (2 p-1)! \zeta(2p) \beta^p + (-1)^{p+1} 2^p
 (2p-2)!  \zeta(2p-1) \frac1{\beta^{p-1}} + O(\beta^{-p})
\ee
as well as the small $\beta$ expansion
\be
\tilde C^{(\beta)}_{2p} \simeq  (-1)^{p+1} 2^{2-p} (2p-2)! \zeta(2p-1) \beta^{p-1}  \quad , \quad \beta \ll 1
\ee

\medskip

%
%
%
As an example we give more explicitly the fourth cumulant
\bea \label{C4ser}
\tilde C^{(\beta)}_4 = - 12 \sum_{\nu=0}^{\infty} \sum_{q=1}^{\infty} \frac{1}{(\nu \sqrt{\frac{\beta}{2}} + q
\sqrt{\frac{2}\beta})^4 } = - \frac{8}{\beta^2} \sum_{q=1}^{\infty} \psi^{(3)}(\frac{2 q}{\beta})
= - \frac{1}{2} \beta^2  \sum_{\nu=0}^{\infty} \psi^{(3)}(1 + \frac{\beta \nu}{2})
\eea
One can then check that this formula, valid for any $\beta$, correctly reproduces for
the cases $\beta=2 s/r$, with $s,r$ mutual primes, the same result as the original formula \eqref{s1}, for instance one finds
\be
\tilde C^{(\beta=2)}_4 = -12 \zeta (3) \quad , \quad \tilde C^{(\beta=1)}_4 = \frac{\pi ^4}{4}-24 \zeta (3)
\quad , \quad \tilde C^{(\beta=4)}_4 = -24 \zeta (3)-\frac{\pi ^4}{4}
\ee
Let us also give more detailed asymptotics at large and small $\beta$
\bea \label{asympt1}
&& \tilde C^{(\beta)}_4 =  -\frac{1}{30} \pi ^4 \beta ^2
-\frac{8 \zeta (3)}{\beta }+\frac{4 \pi ^4}{15 \beta
   ^2}-\frac{32 \zeta (5)}{\beta
   ^3}+O( \beta^{-4} )  \\
&& = -2 \beta  \zeta (3)-\frac{\pi ^4 \beta
   ^2}{60}-\frac{\beta ^3 \zeta (5)}{2}+O\left(\beta
   ^5\right)
\eea
The fourth cumulant is plotted as a function of $\beta$ in the Figure \ref{Fig1},
together with the large and small $\beta$ asymptotics which, as we
see, are quite accurate.

{\bf Second cumulant $\tilde C^{(\beta)}_2$}. The second cumulant reads, for $\beta = 2 s/r$
\be  \label{s4}
\tilde C^{(\beta)}_{2} = - \frac{2}{r s}
\sum_{\nu=0}^{r-1} \sum_{q=1}^{s} \phi_{2}(\frac{\nu}{r}+\frac{q}{s}) - 2 \log r
\ee
To study the limit where both $r,s \to +\infty$ with a fixed (more precisely, converging) ratio $\beta=2 s/r$,
it is useful to decompose $\phi_2(x)= - \frac{1}{x^2} + \tilde \phi_2(x)$, where $\tilde \phi_2(x)$
is regular at $x=0$, and to introduce $\sum_{\nu=0}^{r-1} \frac{1}{1+ \nu} = H_r \simeq
\log r + \gamma_E + O(1/r)$. Then one has in that limit
\be
- \frac{2}{r s}
\sum_{\nu=0}^{r-1} \sum_{q=1}^{s} \tilde \phi_{2}(\frac{\nu}{r}+\frac{q}{s}) \to -2 \int_0^1 dx \int_0^1 dy
\, \tilde \phi_2(x+y) = - 2 ( \int_0^1 ds s \tilde \phi_2(s) + \int_1^2 ds (2-s) \tilde \phi_2(s) )
= 2 \log 2
\ee
Hence need to evaluate the limit
\bea \label{C200}
&& \tilde C^{(\beta)}_{2} \simeq 2 \log 2 +  2 \gamma_E + 2
\sum_{\nu=0}^{r-1} [ \sum_{q=1}^{s} \frac{\beta/2}{(\nu \frac{\beta}{2} + q)^{2} }  - \frac{1}{1+ \nu}  ] \\
&& = 2 \log 2 +  2 \gamma_E + 2
\sum_{\nu=0}^{r-1} [ \frac{\beta}{2} \psi^{(1)}(1 + \frac{\beta \nu}{2}) -
\frac{\beta}{2} \psi^{(1)}(1 + s + \frac{\beta \nu}{2}) - \frac{1}{1+ \nu}  ]
\eea
where we have used that $\sum_{q=1}^s \frac{1}{(q+a)^2} = \psi^{(1)}(1 + a)- \psi^{(1)}(1 + s + a)$.
Now one can check that
\be
\lim_{r \to + \infty} \sum_{\nu=0}^{r-1}  \frac{\beta}{2} \psi^{(1)}(1 + \frac{\beta}{2} r + \frac{\beta \nu}{2})  \simeq
\lim_{r \to + \infty}  \sum_{p=r}^{2 r-1} \frac{1}{\frac{2}{\beta} + p} = \log 2
\ee
where the second line is obtained writing $p=r + \nu$ and using $\psi^{(1)}(x) \sim 1/x$ at large $x$, but
the full equivalence has also been confirmed numerically. Hence we can take the
large $r,s$ limit in \eqref{C200}, the factors $\log 2$ cancel, and we finally obtain the second cumulant for any $\beta$ as the following convergent "dual"
series
\bea \label{C2000}
 \tilde C^{(\beta)}_{2} = 2 \gamma_E  + 2
\sum_{\nu=0}^{+\infty} [ \sum_{q=1}^{+\infty} \frac{\beta/2}{(\nu \frac{\beta}{2} + q)^{2} }  - \frac{1}{1+ \nu}  ]
& = &
2 \gamma_E  + 2
\sum_{\nu=0}^{+\infty} [ \frac{\beta}{2} \psi^{(1)}(1 + \frac{\beta \nu}{2}) - \frac{1}{1+ \nu}  ] \\
& = & 2 \gamma_E  + 2 \log (\beta/2) + 2 \sum_{q=1}^{\infty} ( \frac{2}{\beta} \psi^{(1)}(\frac{2 q}{\beta}) - \frac{1}{q} )
\eea
Note the non trivial term $2 \log(\beta/2)$ in the last expression, arising from the replacement $-2 \log r = -2 \log s + 2 \log (\beta/2)$ in \eqref{s4}. For $\beta=2$ one recovers $\tilde C^{(\beta=2)}_{2}=2 + 2 \gamma_E$. We also
find either from \eqref{C2000}, or from the original formula \eqref{s4}
\be
\tilde C^{(\beta=1)}_{2}=2+2 \gamma_E -\frac{\pi ^2}{4} \quad , \quad
\tilde C_2^{(\beta=4)} = 2+2 \gamma_E +\frac{\pi ^2}{4}+\log (4)
\ee
One obtains the series at large and small $\beta$
\bea \label{asympt2}
&& \tilde C^{(\beta)}_{2} = \frac{\pi ^2 \beta }{6}+2 \gamma_E -\frac{\pi ^2}{3 \beta }+\frac{4 \zeta (3)}{3 \beta
   ^2}-\frac{16 \zeta (5)}{15 \beta ^4}+O(\beta^{-5}) \\
   && = 2 \log (\beta/2)  + 2 \gamma_E
   + \frac{\pi ^2 \beta }{12}+\frac{\beta ^2 \zeta (3)}{12}-\frac{\beta ^4 \zeta
   (5)}{240}+O\left(\beta ^5\right)
\eea
Note that the leading term agrees with \eqref{s4} although that result assumed $p \geq 2$.

The second cumulant is plotted as a function of $\beta$ in the Figure \ref{Fig1}
together with the large and small $\beta$ asymptotics which, as we
see, are again quite accurate.

\begin{figure}[hb]
\includegraphics[width = 0.4\linewidth]{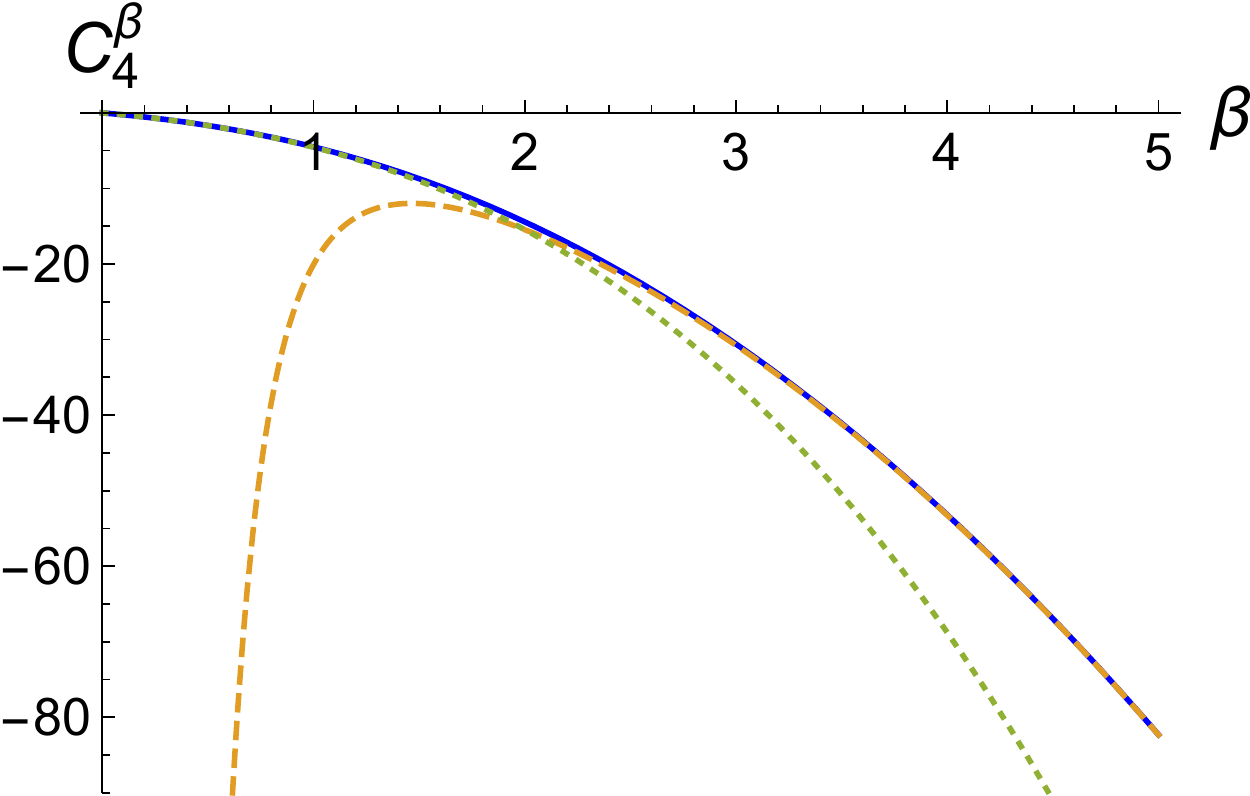}
\includegraphics[width = 0.4\linewidth]{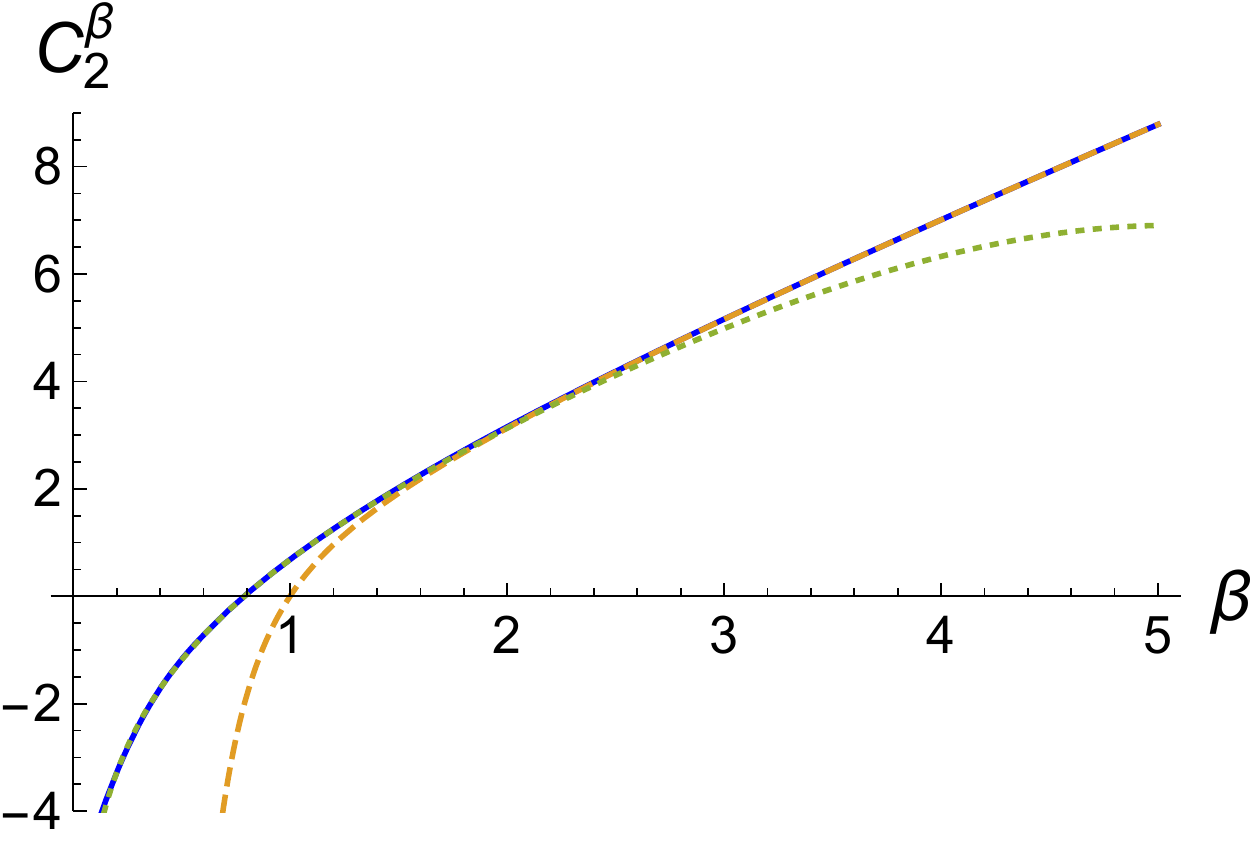}
{\caption{Left: fourth cumulant amplitude $\tilde C^{(\beta)}_{4}$ plotted (in blue)
as a function of $\beta$ from \eqref{C4ser}. Dotted and dashed lines are small and large $\beta$ asymptotics, \eqref{asympt1} respectively.
Right: same for second cumulant amplitude $\tilde C^{(\beta)}_{2}$ (in blue) from
\eqref{C2000} and \eqref{asympt2}.}\label{Fig1}}
\end{figure}

\section{Cumulant amplitudes $C_k(\ell)$}

Let us recall the result given in the text for the amplitudes $C_k(\ell)$ for
$\ell=2 \pi$ and $\ell \ll 1$. For any $k \geq 2$
\bea \label{restot}
&& C_k(2 \pi) = (-1)^k \frac{d^k}{dt^k}|_{t=0}  \log \bigg[
 \frac{\Gamma(1+ t )^2 G(2-2 t)}{G(2-t)^3 G(2+t)} \bigg] \\
&& C_k(\ell) \simeq 2 \log \ell \, \delta_{k,2} + (-1)^k \frac{d^k}{dt^k}|_{t=0} \bigg[
\frac{2  \Gamma(1+ t )^2  G(2-2 t)}{G(2+t)^2  G(2-t)
G(4-t)} \bigg]
\eea
We now use Eq. \eqref{eqG}, and we also use that $\psi^{(k)}(1)=(-1)^{k+1} k! \zeta(k+1)$ and
$\psi^{(k)}(2)=\psi^{(k)}(1) + (-1)^{k} k!$ and  $\psi^{(k)}(4)=\psi^{(k)}(1) + (-1)^{k} k!(1+ 2^{-k-1} + 3^{-k-1})$. We obtain for the full circle
\bea
&& C_2(2 \pi) = \frac{\pi^2}{3} \quad , \quad C_3(2 \pi) = 2 \pi^2 - 8 \zeta(3)  \quad , \quad C_4(2 \pi) = \frac{14}{15} \pi^4 - 72 \zeta(3) \\
&&
C_k(2 \pi) = \left(\left(1-(-2)^k+3 (-1)^k\right) \zeta
   (k-1)+\left(1+(-2)^k-3 (-1)^k\right) \zeta
   (k)\right) \Gamma (k) \quad , \quad k \geq 3 \nn
\eea
and for the mesoscopic interval $\ell \ll 1$
\bea
&& C_2(\ell)= 2 \log \ell + \frac{9}{4}
\quad , \quad C_3(\ell) = - \frac{17}{4} + \frac{8 \pi^2}{3} - 8 \zeta(3)  \quad , \quad
C_4(\ell) =\frac{99}{8} +  \frac{4}{5} \pi^4 - 72 \zeta(3)
\\
&&
C_k(\ell) =
6^{-k} \left(\left(-(-12)^k+(-3)^k 2^{k+1}+2^{k+1}
   3^k\right) \zeta (k-1)+(-6)^k \left(2^k-4\right)
   \zeta (k)+(-3)^k \left(2^{k+1}+1\right)\right)
   \Gamma (k) ~,~ k \geq 3 \nn
\eea

\section{{Distribution of the maximum over a mesoscopic interval $\frac{1}{N} \ll \ell \ll 2 \pi$}}
Here we sketch the derivation of our second main result,
 \eqref{ltinterval}.

To this end we set $\phi_a = \theta_A + \ell x_a$, with
$x_a \in [0,1]$, and recall $\ell=\theta_B-\theta_A$.
 Eq \eqref{Zn}
gives
\bea \label{Zn2}
&& \mathbb{E}[Z_b^n]  \simeq
\left(\frac{N \ell}{2 \pi}\right)^n  (N \ell)^{b^2 (n + n^2) }
|A_{\beta}(b)|^{2n} |A_{\beta}(b n)|^{2}  \prod_{a=1}^n \int_{0}^{1} d x_a
\prod_{1 \leq a < c \leq n}
|x_a-x_c|^{- 2 b^2}
\prod_{1 \leq a \leq n}
x_a^{2 n b^2}
\eea
One recognizes now the Selberg integral \cite{ForresterWarnaar} in the form
which arises in the study of the fBm0 on an interval
\cite{FLD2016,CaoPathologies}. Using the known expression for its analytical
continuation (see Eq. (239) in \cite{FLD2016}) and following the
same steps as for the full circle presented in the text, we obtain the DSLT in the high temperature phase $b<1$
with $t=-n b$ as \cite{footnoteF}
\bea
&& \mathbb{E} \left(e^{- 2 \pi\sqrt{\frac{\beta}{2}} \left({\cal F}-\tilde{{\cal F}}_1\right) t} \right) \simeq  (N \ell)^{- t Q + t^2}
|A_{\beta}(-t)|^{2}  \Gamma(1+ t b)
\frac{G_b(Q)^2 G_b(Q-2 t)
G_b(2Q)}{G_b(Q+t)^2  G_b(Q-t)
G_b(2 Q-t)}
  \label{eq10}
\eea
The duality invariance can be similarly checked and from the FDC we find that
the DSLT in the low temperature phase $b>1$
is given again by \eqref{rel} with our second
main result, i.e. Eq. \eqref{ltinterval} of the main text.

\section{Verification of relation eq.(\ref{eq5})}

Our starting point is the characteristic polynomial defined in the text, which we rewrite
\[
\xi_N(\theta) = \prod_j\left( 1 - e^{ i (\theta_j-\theta)} \right)=\prod_j e^{ i (\theta_j-\theta-\pi)/2}\, 2\sin{\frac{\theta_j-\theta}{2}}
\]
\be \label{xiSM}
=e^{-i\frac{N(\pi+\theta)}{2}+\frac{i}{2}\sum_{j=1}^N\theta_j}\prod_{j=1}^N\,
\mbox{sgn}\left[\sin{\frac{\theta_j-\theta}{2}}\right]\prod_{j=1}^N
2\left|\sin{\frac{\theta_j-\theta}{2}}\right|
\ee
Further using
\[
\prod_{j=1}^N\,
\mbox{sgn}\left[\sin{\frac{\theta_j-\theta}{2}}\right]=\prod_{j=1}^N\,
\mbox{sgn}(\theta_j-\theta)=(-1)^{\#(\theta_j<\theta)}=e^{i\pi \#(\theta_j<\theta)}
\]
where $\#(\theta_j<\theta):=\sum_{j=1}^N \chi(\theta - \theta_j)$ is the number of $\theta_j$ not exceeding the value $\theta$, we see that
\be
 {\rm Im} \log \xi_N(\theta) =-\frac{N}{2}(\pi+\theta)+\frac{1}{2}\sum_{j=1}\theta_j+\pi\#(\theta_j<\theta)
\ee
implying via the definition (\ref{defcount})
\be
\frac{1}{\pi} {\rm Im} \log \xi_N(\theta) -\frac{1}{\pi} {\rm Im} \log \xi_N(\theta_A)=-\frac{N}{2\pi}(\theta-\theta_A)+
\#(\theta_j<\theta)-\#(\theta_j<\theta_A):=\delta {\cal N}_{\theta_A}(\theta)
\ee
 exactly as claimed in (\ref{eq5}).

\section{Deficiency of the log-correlated Gaussian approximation for characterising the maximum of $\delta {\cal N}_{\theta_A}(\theta)$.}
Let us demonstrate that naively replacing the difference $\delta {\cal N}_{\theta_A}(\theta)$ with its Gaussian approximation $\frac{1}{\pi}\left[ W_{\beta}(\theta)-W_{\beta}(\theta_A)\right]$   is not sufficient for the purpose of characterizing the maximum of the process. For this end, we make the corresponding replacement
in the expression  (first line of \ref{15}) for the integer moments of $Z_b$, yielding
\be
\mathbb{E}[Z_b^n]=\left(\frac{N}{2 \pi}\right)^n   \int_{\theta_A}^{\theta_B}
\, \mathbb{E} \left[ e^{2  b \sqrt{\beta/2} \left(\sum_{a=1}^n W_{\beta}(\phi_a)-nW_{\beta}(\theta_A)\right)} \right]\,  \prod_{a=1}^n d \phi_a \label{firstline}
\ee
where now the expectation is over the mean-zero Gaussian process $W_{\beta}(\theta)$ with the covariance given by
(\ref{covcharpol}) and the variance $\mathbb{E}( W(\theta)^2)=\beta^{-1}\log{N}$. Due to Gaussian nature of the process the expectation is readily taken via the identity
\[\mathbb{E} \left[ e^{2  b \sqrt{\beta/2} \left(\sum_{a=1}^n  W_{\beta}(\phi_a)-nW_{\beta}(\theta_A)\right)} \right]=
e^{\beta b^2\left[n(n+1)\mathbb{E}\left( W(\theta_A)^2\right)+2\sum_{a<c}^n\mathbb{E}\left( W(\phi_a)W(\phi_c)\right) -2n\sum_{a=1}^n\mathbb{E}\left( W(\phi_a) W(\theta_A)\right)\right] }
\]
Substituting here the value
(\ref{covcharpol}) for the covariance and the associated variance
and recalling that $4 \sin^2\left(\frac{\theta_1-\theta_2}{2}\right)=|e^{i\theta_1}-e^{i\theta_2}|^2$
we immediately arrive at the expression for the moments:
\bea
&& \mathbb{E}[Z_b^n]  \simeq
\left(\frac{N}{2 \pi}\right)^n  N^{b^2 n(n+1) }
  \int_{\theta_A}^{\theta_B}
\prod_{1 \leq a < c \leq n}
|1- e^{i (\phi_a-\phi_c)}|^{- 2 b^2} \\
&&
\times \prod_{1 \leq a \leq n}
|1- e^{i (\phi_a- \theta_A)}|^{2 n b^2}\, \prod_{a=1}^n d \phi_a \nn
\eea
which misses exactly the factor $|A_{\beta}(b)|^{2n} |A_{\beta}(b n)|^{2}$ when compared to the formula (\ref{momasy}). As those factors contribute to the cumulants for the maximum of the process, the Gaussian approximation is clearly insufficient for this purpose.

\section{DSLT in the complex plane}

The formulas \eqref{eq1}, \eqref{ltcircle}, \eqref{ltinterval}, of the main text 
for the DSLT
were obtained for real values of the parameter $t$. We expect
them to extend to a domain around $t=0$ in the complex plane.
For real $t$ this domain cannot contain $t=Q/2$ for $b>1$
and $t=1$ for $b \leq 1$, which is the location of a termination point transition
for the pinned fBm0 (it corresponds to events when the minimum is
at $\theta_m \approx \theta_A$), analyzed in \cite{CaoPathologies,CaoOPELiouville}.
The domain should also be contained within ${\rm Im}(t) < 1/2$ (for $\beta=2$) because
of the integer nature of the field ${\cal N}(\theta)$.
Extending the formula beyond remains open.
However, results from \cite{DeiftItsKrasovsky2009,AbanovIvanovQian2011} for
$\beta=2$, suggest that, treating $n b=-t$ and $b$ as
independent variables, the integrand in \eqref{Zn} can be extended
formally to a sum over $b n \to b n + i \mathbb{Z}$, $b \to b + i \mathbb{Z}$.
Investigating these properties is left for future studies.

\section{Moments of the position of the maximum}

As to the position of the maximum  $\theta_m \in [\theta_A,\theta_B]$,
we recall that its statistics for the fBm0 on an interval has been investigated in \cite{FLD2016}
by calculating those of
the Jacobi ensemble of random matrices and performing the continuation to $n=0$.
Defining
$y_m = (\theta_m - \theta_A)/\ell$, the moments $\mathbb{E}(y_m^k)$
are thus the ones given in \cite{FLD2016} (in Eqs. (129-130) for $k=2,4$
and Eqs. (101),(98-100), and Appendix C for general $k$).
Extending that calculation to treat the case of the mesoscopic interval,
one checks that the additional factors in \eqref{Zn2} do not contribute,
and arrives at the results mentioned in the Letter.

\section{Joint moments of the position and the value of the maximum}

{\it Preliminary remark}. Consider two random variables $X_1$ and $X_2$. By definition the connected moments (also called bivariate cumulants) are given by
\be \label{multi}
\mathbb{E}_c(X_1^{q_1} X_2^{q_2}) = \partial^{q_1}_{t_1}|_{t_1=0} \partial^{q_2}_{t_2}|_{t_2=0} \log \mathbb{E}(e^{t_1 X_1 + t_2 X_2})
\ee
Let us define the following biased average
\be
\langle f(X_2) \rangle_{t_1}=
\frac{\mathbb{E}(f(X_2) e^{t_1 X_1})}{\mathbb{E}(e^{t_1 X_1})}
\ee
Expanding the r.h.s of \eqref{multi} in powers of $t_2$ we see that
\be
 \mathbb{E}_c(X_2 X_1^{q}) = \partial^{q}_{t_1}|_{t_1=0} \langle X_2 \rangle_{t_1} \quad , \quad   \mathbb{E}_c(X_2^2 X_1^{q}) = \partial^{q}_{t_1}|_{t_1=0} (\langle X_2^2 \rangle_{t_1} -
\langle X_2 \rangle_{t_1}^2)
\ee
and so on, which is also equivalent (upon multiplying by $1/q!$ and summing over $q$) to
the following formula for the generating functions of the bivariate cumulants of lowest
order in $X_2$
\be
 \mathbb{E}_c(X_2 e^{t_1 X_1}) = \langle X_2 \rangle_{t_1}  = \frac{\mathbb{E}(X_2 e^{t_1 X_1})}{\mathbb{E}(e^{t_1 X_1})}
 \quad , \quad
  \mathbb{E}_c(X_2^2 e^{t_1 X_1}) = \langle X_2^2 \rangle_{t_1} -
\langle X_2 \rangle_{t_1}^2
  \ee
which will be useful below.
%

\subsection{Mesoscopic interval}

Let us discuss first the mesoscopic interval.
Let us denote, as in the text, $y= \frac{\theta - \theta_A}{\ell} \in [0,1]$,
and $ \langle y^k \rangle$ the $k$-th moment of the random variable $y$ with respect to
the Gibbs measure associated to $Z_b$ defined in \eqref{Zb}, for a fixed random
configuration of the eigenvalues $\theta_i$. One can evaluate the following ratio of averages w.r.t.
the measure CUE$_\beta$ for the eigenvalues
\bea \label{fr}
\frac{ \mathbb{E}[ \langle y^k \rangle Z_b^n ] }{\mathbb{E}[Z_b^n ]} = \langle y^k \rangle_{\beta,a,b,n}|_{(\beta,a,b)\to (b,2 n b^2, 0)} = M_k(t=-bn, b)
\eea
The numerator in the l.h.s. of \eqref{fr} equals Eq. \eqref{Zn2} of the text with $x_a \to y_a$
and $y_1^k$ inserted in the integrand. The corresponding ratio is thus
the $k$-th moment of the Jacobi ensemble denoted $\langle y^k \rangle_{\beta,a,b,n}$
in \cite{FLD2016} with the identification of parameters corresponding to fBm0
(see Eqs. (56,57) there). Note that the extra factors containing $A_\beta(z)$ in
\eqref{Zn2}, not present in the fBm0, drop out in the ratio. The expression
for the $\langle y^k \rangle_{\beta,a,b,n}$ were obtained in \cite{FLD2016} and
we denote $M_k(t=-bn, b)$ these expressions, which are rational fractions of
the variables $t=-b n$ and $b$. We thus obtain
\be \label{qq}
\mathbb{E}( \langle y^k \rangle e^{- 2 \pi \sqrt{\frac{\beta}{2} } t {\cal F}} ) = M_k(t,b) \mathbb{E}(e^{- 2 \pi \sqrt{\frac{\beta}{2} } t {\cal F}} )
\ee
which is valid in the high temperature phase $b<1$.
The simplest examples are the first two moments $k=1,2$. From (107) and (190) in \cite{FLD2016} we obtain
\be
M_1(t,b) = \frac{1}{2} - \frac{t b}{2 (1+ b^2)} \quad , \quad
M_2(t,b) = \frac{\left(b^2-b t+1\right) \left(b \left(b \left(4
   b^2-b t+t^2+9\right)-t\right)+4\right)}{2 \left(6
   b^6+19 b^4+19 b^2+6\right)}
\ee
These expressions are duality invariant, i.e. does not change under $b \to 1/b$.
All moments
$M_k(t,b)$ share this property \cite{FLD2016}
(their explicit expressions are given in (91-92) there).
Hence the freezing duality conjecture (FDC) allows to continue \eqref{qq} for $b>1$. As in the text, the r.h.s. is duality invariant if
multiplied by $\Gamma(1+ \frac{t}{b})$, hence the value of the l.h.s, as a function of $b$, freezes at $b=1$.
Taking $b \to +\infty$ we obtain
\be \label{kkk1}
\mathbb{E}(y_m^k e^{- 2 \pi \sqrt{\frac{\beta}{2} } t \delta {\cal N}_m} ) = M_k(t,1) \mathbb{E}(e^{- 2 \pi \sqrt{\frac{\beta}{2} } t \delta {\cal N}_m} )
\ee
where $y_m$ is the position of the maximum. Setting $t=0$ yields the results for the moments
$\mathbb{E}(y_m^k)$ quoted in the text. Let us denote the centered variables
\be
\tilde y_m = y_m - \mathbb{E}(y_m) \quad , \quad
\delta \tilde {\cal N}_m=
\delta {\cal N}_m-\mathbb{E}(\delta {\cal N}_m)
\ee
Consider \eqref{kkk1} for $k=1$. Using that $\mathbb{E}(y_m)
=\frac{1}{2}$, this can be written as
\be \label{a11}
\mathbb{E}( \tilde y_m  e^{- 2 \pi \sqrt{\frac{\beta}{2} } t
\delta \tilde {\cal N}_m
} ) = - \frac{t}{4}   \, \mathbb{E}(e^{- 2 \pi \sqrt{\frac{\beta}{2} } t \delta \tilde {\cal N}_m} ) \quad  \Leftrightarrow
\quad \langle \tilde y_m \rangle_t = - \frac{t}{4}
\ee
where it is useful to define the following averages
\be
\langle f(\tilde y_m) \rangle_t := \frac{\mathbb{E}( f(\tilde y_m)  e^{- 2 \pi \sqrt{\frac{\beta}{2} } t
\delta \tilde {\cal N}_m} ) }{\mathbb{E}(e^{- 2 \pi \sqrt{\frac{\beta}{2} } t \delta \tilde {\cal N}_m} )}
\ee
which represent averages under a biased probability
$e^{- 2 \pi \sqrt{\frac{\beta}{2} } t \delta \tilde {\cal N}_m}  \times
 {\cal P}(\delta \tilde {\cal N}_m)$, where ${\cal P}$ is the
PDF of $\delta \tilde {\cal N}_m$. Here $t<0$ corresponds to biasing the values of
the maximum towards the positive values and leads to positive values
on average for $\tilde y_m$. Expansion in powers of $t$ in \eqref{a11} yields the relations, valid for any $p \geq 0$
\be \label{a2}
\mathbb{E}\big(\tilde y_m   (\delta \tilde {\cal N}_m)^p \big)
= \frac{p}{4} \frac{1}{2 \pi \sqrt{\beta/2}} \mathbb{E}\big( (\delta \tilde {\cal N}_m)^{p-1} \big)
= p \, \mathbb{E}\big(\tilde y_m  \delta \tilde {\cal N}_m  \big) \times
\mathbb{E}\big( (\delta \tilde {\cal N}_m)^{p-1} \big)
\ee
In particular
\be \label{a3}
\mathbb{E}\big( \tilde y_m  (\delta \tilde {\cal N}_m)^p \big)
= \frac{1}{(2 \pi \sqrt{\beta/2})^p} \times  \begin{cases}
\frac{1}{4} \quad , \quad p=1 \\
0  \quad , \quad p=2 \\
\frac{3}{4} (2 \log N \ell + \frac{9}{4}) \quad , \quad p=3 \\
(- \frac{17}{4} + \frac{8 \pi^2}{3} - 8 \zeta(3) ) \quad , \quad p=4
\end{cases}
\ee
The result for $p=1$ shows that positions of maximum $y_m>1/2$ correlate with
values of the maximum larger than the average, consistent with the pinning at $y=0$, i.e.
$\delta {\cal N}(\theta=\theta_A)=0$,
while the boundary condition at $y=1$ is free. Since $\delta {\cal N}_m-\mathbb{E}(\delta {\cal N}_m)$
is typically $\sim \sqrt{\log N \ell}$ the correlation with the Gaussian part of the fluctuations of
the value of the maximum is absent in the correlation for $p=1$ (which is $O(1)$). Now, it is easy to see
that \eqref{a2} and \eqref{a11} imply that {\it all higher bi-variate cumulants vanish}, i.e.
the information contained in \eqref{a11} can be summarized as
\bea
&& \mathbb{E}\big( \tilde y_m  \delta \tilde {\cal N}_m \big)
= \frac{1}{4} \frac{1}{2 \pi \sqrt{\beta/2}} \\
&&  \mathbb{E}_c\big( \tilde y_m  (\delta \tilde {\cal N}_m)^p \big) = 0  \quad , \quad p \geq 2
\eea
consistent with \eqref{a3} which is the sum of all {\it disconnected} averages.

For $k=2$ we obtain
\be \label{a1}
\mathbb{E}(y_m^2  e^{- 2 \pi \sqrt{\frac{\beta}{2} } t
\delta \tilde {\cal N}_m
} ) =  \frac{1}{100} (2-t) \left(17 - 2 t + t^2\right)  \, \mathbb{E}(e^{- 2 \pi \sqrt{\frac{\beta}{2} } t \delta \tilde {\cal N}_m} )
\quad  \Leftrightarrow
\quad \langle y_m^2 \rangle_t = \frac{1}{100} (2-t) \left(17 - 2 t + t^2\right)
\ee
For $t=0$ we obtain the result given in the text
$\mathbb{E}(y_m^2)=\frac{17}{50}$. Expansion of the first equation in powers of
$t$ allows to obtain all joint moments of the form $\mathbb{E}(y_m^2 (\delta \tilde {\cal N}_m)^p )$ using our results for the cumulants of the value of the maximum (given in the text).
Alternatively we may write the bi-variate cumulants (see preliminary remark above)
\be
\mathbb{E}_c(y_m^2  e^{- 2 \pi \sqrt{\frac{\beta}{2} } t \delta \tilde {\cal N}_m} )
= \langle y_m^2 \rangle_t - \langle y_m \rangle_t^2 = \frac{1}{400} (4-t^2) (9+4 t)
\ee
Expanding in powers of $t$ on both sides we see that for $k=2$, only the first three connected
moments are non zero.

\subsection{Full circle}

Consider now the average of $\cos k \phi$ with respect to the Gibbs measure associated to $Z_b$,
defined in \eqref{Zb}, on the full circle, i.e. with $\theta_A=-\pi$ and $\theta_B=\pi$. Again one
can evaluate the ratio of averages w.r.t.
the measure CUE$_\beta$ for the eigenvalues
\bea \label{fr2}
\frac{ \mathbb{E}[ \langle \cos k \phi \rangle Z_b^n ] }{\mathbb{E}[Z_b^n ]} =
\langle \cos(k \phi) \rangle_{\beta,\mu,n}
= (-1)^k \langle y^k \rangle_{\beta,a,b,n}|_{(\beta,a,b)\to (b,-1-b^2,2 n b^2)} = \tilde M_k(t=-bn, b)
\eea
The numerator in the l.h.s. of \eqref{fr2} equals Eq. \eqref{Zn} of the text with
$\cos k \phi_1$ inserted in the integrand. The second equality is the conjecture obtained in
Eqs. (157-158) in \cite{FLD2016} (where $\langle \cos(k \phi) \rangle_{\beta,\mu,n}$
denotes the l.h.s. in (158) there, with $\kappa=-\beta^2 \to - b^2$ and $\mu \to n b^2$)
which relates the moments in the circular ensemble to those on the interval, i.e the moments
$\langle y^k \rangle_{\beta,a,b,n}$ already used in the previous section. The different
specialisations of the parameters leads to other rational functions $\tilde M_k(t=-bn, b)$.
From (107) and (91-92) in \cite{FLD2016} we obtain
\be
\tilde M_1(t,b) = -  \frac{t b}{1+ b^2} \quad , \quad
\tilde M_2(t,b) = \frac{b t \left(b^4+2 b^3 t-b^2 t^2+3 b^2+2 b t+1\right)}{\left(b^2+1\right)
   \left(b^2+2\right) \left(2 b^2+1\right)}
\ee
These expressions are again duality invariant. From the FDC we obtain
\be \label{kkk}
\mathbb{E}(\cos(k \theta_m) e^{- 2 \pi \sqrt{\frac{\beta}{2} } t \delta {\cal N}_m} ) = \tilde M_k(t,1) \mathbb{E}(e^{- 2 \pi \sqrt{\frac{\beta}{2} } t \delta {\cal N}_m} )
\ee
where $\theta_m$ is the position of the maximum of $\delta {\cal N}(\theta)$ on the full circle. We can check that for $t=0$ all
$\mathbb{E}(\cos(k \theta_m))=0$, $\theta_m$ has a uniform distribution on the circle.
For $k=1$ we obtain
\be
\mathbb{E}(\cos(\theta_m)  \delta \tilde {\cal N}_m ) =  \frac{1}{2} \frac{1}{2 \pi \sqrt{\beta/2}}
\ee
which shows that higher values of the maximum correlate with $\cos(\theta_m) >0$,
i.e. $\theta_m$ being closer to $0$ than to $\pm \pi$, the point at which its value is pinned to zero.
Again we see that all connected correlations $\mathbb{E}_c(\cos(\theta_m)  (\delta \tilde {\cal N}_m)^p )$
vanish. For $k=2$ we obtain
\be \label{a6}
\mathbb{E}(\cos(2 \theta_m)   e^{- 2 \pi \sqrt{\frac{\beta}{2} } t
\delta \tilde {\cal N}_m
} ) =  \frac{1}{18} (5-t) t (t+1) \, \mathbb{E}(e^{- 2 \pi \sqrt{\frac{\beta}{2} } t \delta \tilde {\cal N}_m} )
\ee
which, upon expanding in $t$ yields all joint moments of the form $\mathbb{E}(\cos(2 \theta_m)
(\delta \tilde {\cal N}_m)^p)$. \\

\end{widetext}


\begin{thebibliography}{99}
\bibitem{footnote2}  To understand this, consider a discrete pinned field on the circle constructed
as $V_i=U_i-U_1$,  $i=1,\dots M$ where $U_i$ is a standard centered log-correlated
field. Clearly the position of the maximum for $U_i$, hence of the one for $V_i$,
is uniformly distributed on the circle.
\bibitem{footnoteF} We defined ${\cal F}_1=\frac{1}{\pi b \sqrt{2\beta}}\log{K_{\beta}(b)}$ with $K_{\beta}(b)=|A_{\beta}(b)|^2/\left(2 \pi \Gamma(1-b^2)\right)$ and similarly for $\tilde{{\cal F}}_1$ with $\tilde K_{\beta}(b)=K_{\beta}(b) (2 \pi)^{\beta-1}$
\bibitem{footnoteComplex}
The DSLT for complex $t$ is discussed in
\cite{SM}.
\end{thebibliography}
\end{document}